\documentclass[journal=jpccck,manuscript=article]{achemso}

\usepackage{amsmath}

\author{Giovanni Pireddu}
\affiliation{Dipartimento di Chimica e Farmacia, Universit\`a degli Studi di Sassari, Via Vienna 2, 01700 Sassari, Italy}
\email{gpireddu@uniss.it}
\author{Federico G. Pazzona}
\affiliation{Dipartimento di Chimica e Farmacia, Universit\`a degli Studi di Sassari, Via Vienna 2, 01700 Sassari, Italy}
\author{Alberto M. Pintus}
\affiliation{Dipartimento di Chimica e Farmacia, Universit\`a degli Studi di Sassari, Via Vienna 2, 01700 Sassari, Italy}
\author{Andrea Gabrieli}
\affiliation{Dipartimento di Chimica e Farmacia, Universit\`a degli Studi di Sassari, Via Vienna 2, 01700 Sassari, Italy}
\author{Pierfranco Demontis}
\affiliation{Dipartimento di Chimica e Farmacia, Universit\`a degli Studi di Sassari, Via Vienna 2, 01700 Sassari, Italy}

\title{Spatial coarse-graining of methane adsorption in graphene materials}

\begin{document}


\begin{abstract}
We investigate the spatial coarse-graining of interactions in host-guest systems within the framework of the recently proposed Interacting Pair Approximation (IPA) [Pazzona~\emph{et al.}, \emph{J.~Chem.~Phys.} \textbf{2018}, \emph{148}, 194108]. 
Basically, the IPA method derives local effective interactions from the knowledge of the bivariate histograms of the number of sorbate molecules (occupancy) in a pair of neighboring subvolumes, taken at different values of the chemical potential.
Here we extend the IPA approach to the case in which every subvolume is surrounded by \emph{more than one class of neighbors}, and we apply it on two systems: methane on a single graphene layer and methane between two graphene layers, at several temperatures and sorbate densities. 
We obtain coarse-grained (CG) adsorption isotherms and reduced variances of the occupancy in a quantitative agreement with reference atomistic simulations. A quantitative matching is also obtained for the occupancy correlations between neighboring subvolumes, apart from the case of high sorbate densities at low temperature, where the matching is refined by pre-processing the histograms through a quantized bivariate Gaussian distribution model.
\end{abstract}

\pagebreak

\section{Introduction}

The representation of physicochemical phenomena involving molecular systems in a variety of spatial and temporal scales has always been a challenging task.
Nowadays, atomistic computational methods such as \emph{ab-initio} molecular dynamics, offer a very detailed and accurate framework for the study of molecular systems.~\cite{Parrinello1985} 
However, the simulation of relatively large environments requires a considerable computational effort. 
Even with atomistic classical molecular dynamics (MD) and Monte Carlo (MC) methods, the simulation of systems at the meso- and macroscopic scales remains unfeasible. 

This makes the development of coarse-graining protocols an active line of research.
With a possible slight loss of accuracy, the production of less-detailed but more computationally efficient models allows switching from a fine-grained (FG) to a coarse-grained (CG) representation of the system under investigation.
In this line of work we think of such CG description in terms of occupancy-based models of adsorption, where an effective interaction field is defined over the \emph{local occupancy} (that is the number of guest molecules' centers) in the nearness of discrete locations inside the adsorbent rather than on fine-grained atomistic configurations.\cite{vanTassel1993,Saravanan1998,Czaplewski1999,TuncaFord1999,Tunca2004,Demontis1997_JCPB,Auerbach2004,Das2009}

Thus, the coarse-graining approach we follow is of a \emph{spatial} rather than \emph{topological} kind; that is, instead of building CG units out of groups of atoms through mapping operators (which is, in a \emph{very} few words, the spirit of topological coarse-graining~\cite{Bilionis2013,Izvekov2005,Izvekov2005b,Noid2008,Reith2003,Tsourtis2017}), we turn our attention to the partitioning of the system domain in non-overlapping subvolumes and the association of proper CG state variables to each of them.\cite{Ma1976,Katsoulakis2003,Dai2008,Israeli2006,Ayappa1999,Saravanan1998,Snurr1994,Pazzona2013,Pazzona2014,Pazzona2018}
In general, the idea of representing adsorption phenomena through a real-space lattice model is at least one century-old \cite{Langmuir1918}, but methods are still under continuous development, due to the lack of a sufficiently general and accurate protocol.~\cite{Guo2016,Tarasenko2019,Sudibandriyo2010}

Local occupancies are precisely the CG state variables we are focusing on here, and we represent them as discrete stochastic variables. The subvolumes we consider are of nanometer size and above, thus making the resulting CG model a \emph{mesoscopic} model, and we evaluate the matching between the CG and FG representation in terms of statistical properties of occupancy distributions, while neglecting any detail of the original system below that scale. Our study then is aimed to define, at constant temperature, the \emph{effective interactions} between neighboring subvolumes in terms of local occupancies only, within a wide overall density range.
Our effort points towards the development of a general procedure for performing a bottom-up spatial CG of adsorption phenomena while guaranteeing a sufficiently accurate representation of static properties.\\

In our previous paper~\cite{Pazzona2018} we worked on host-guest systems where the neighbors of each adsorption unit (e.g., every $\alpha$-cage of LTA-type zeolites) were all equivalent.
Here, we extend our reasoning to the case where each subvolume is surrounded by neighborhoods of \emph{two} kinds, by making reference to two systems that can be partitioned into two-dimensional square lattices: united-atom methane adsorbed (i) on a single graphene sheet, and (ii) between two graphene sheets.
The latter system is inspired by graphene-based layered materials, which can exhibit interesting properties for the adsorption of chemical species such as methane.~\cite{Yang2018,Hassani2015,Pedrielli2018}

The rest of the paper is organized as follows: we first describe the structure of the CG model and define the relation between CG interaction parameters and occupancy distributions; then we introduce a pre-processing technique that can be used to improve, at low temperature, the agreement between CG and FG adsorption properties; finally, we apply the method to the aforementioned graphene systems, we discuss the results and draw our conclusions. 

\section{Coarse-grained model}\label{sec:model}

\begin{figure}[ht]
  \includegraphics[width=3.3in]{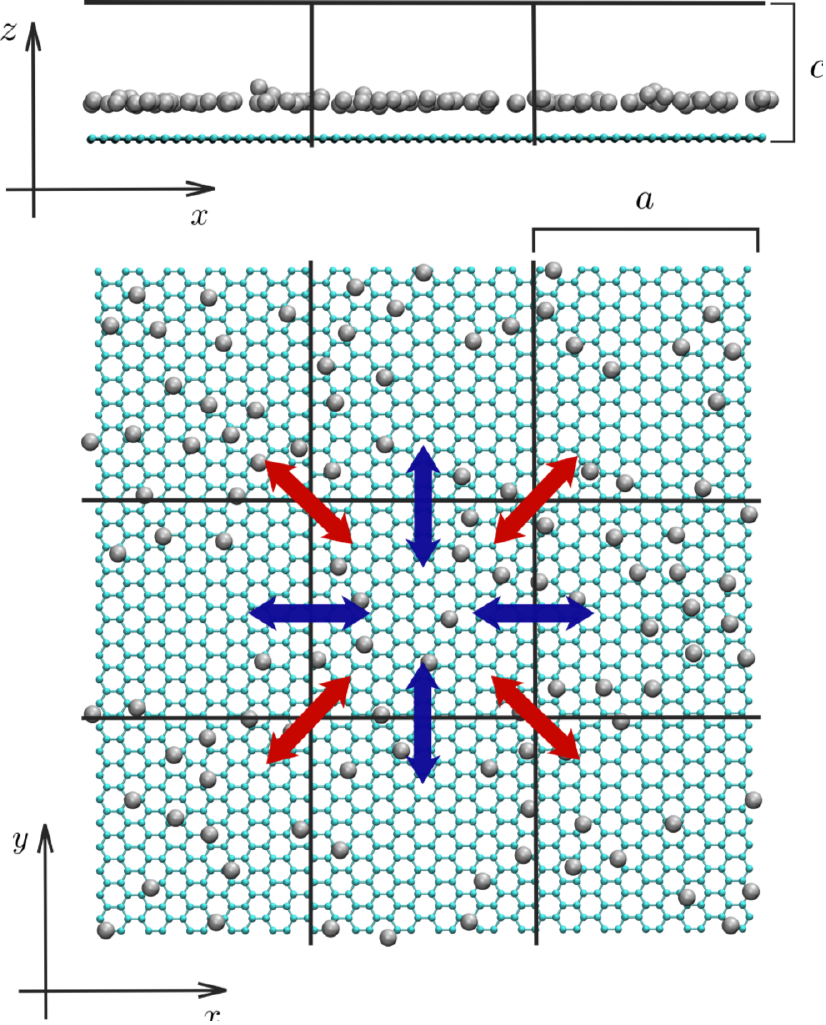}
  \caption{\footnotesize{Two projections of the same snapshot from the FG simulation of the methane and single layer graphene system at $200$ K. In both images the cell partitioning is represented by solid black lines. The bottom image also shows the neighboring classes for the central cell: blue arrows represent class I connections and red arrows represent class II connections.}}
  \label{fgr:Snap}
\end{figure}
In Fig.~\ref{fgr:Snap} we report a picture of a portion of the simulation space of one of our FG systems of interest: a graphene layer (the host) with united-atom methane molecules adsorbed on it (the guests).
As sketched in Fig.~\ref{fgr:Snap}, the space is tessellated with identical, non-overlapping square subvolumes, called \emph{cells}, of edge length $a$.
We say that two cells are neighbors of one another if they share either one edge (\emph{class I neighbors}, center-to-center distance is equal to $a$) or one corner (\emph{class II neighbors}, distance $a\sqrt{2}$).
Therefore, each cell turns out to be connected to $\nu^{\text{I}}=4$ cells of class I, and $\nu^{\text{II}}=4$ cells of class II. The total number of neighbors is denoted as $\nu=\nu^{\text{I}}+\nu^{\text{II}}=8$---one can naturally extend this reasoning to an arbitrary number of neighborhood classes, $\chi=$ I, II, III, \dots, with $\nu= \nu^{\text{I}} + \nu^{\text{II}} + \nu^{\text{III}} + \dots$.
By setting $a=r_c$, where $r_c$ is the cutoff radius used for the potential energy evaluation in the FG simulations, we ensure that no guest molecule in any cell will interact directly with any other molecule outside the neighborhood of that cell.
For any configuration of guest molecules in the space domain, we can count how many of their centers-of-mass fall within every cell; if we label the cells as $i=1,\dots,M$, with $M$ as the total number of cells, the array of integer numbers that results from this counting operation is termed the \emph{occupancy configuration} of the system, and is denoted as $\mathbf{n}=\{n_1,\dots,n_M\}$.
Effective interactions arise inside every cell and between neighboring cells, and neighboring cells of every one class contribute differently to the total effective interaction---this can be easily seen if we think of such interactions in terms of \emph{average}, effective interactions between the $n_i$ particles in cell $i$ and the $n_j$ particles in cell $j$: \emph{on average}, the molecules in a cell will ``feel'' the molecules in the neighborhood of one kind differently from how they ``feel'' those in the neighborhood of another kind.
We consider the system in the grand-canonical ensemble, which is the most common statistical ensemble used to represent adsorption phenomena. In this ensemble, the chemical potential, $\mu$, of the guest species is held constant (along with the temperature $T$), while the overall density fluctuates around the corresponding equilibrium value. 
Due to guest-guest and host-guest interactions (defined on the molecular scale), any change in $\mu$ will cause the properties of the distribution of occupancies in the system to change as well; our aim is to provide our CG square cells with a set of effective, local occupancy-dependent interactions such as to produce (approximately) the same change in the distribution properties.
\\

We define $\Omega$, the CG potential function of the system in the grand-canonical ensemble, as a function of $\mu$ and of its occupancy configuration in the lattice:
\begin{equation}
    \Omega_\mu(\mathbf{n})= 
    \sum_i \left(H_{n_i} - \mu n_i\right)
    + \sum_{\langle ij \rangle} K^{\chi_{ij}}_{n_i,n_j} , 
    \label{eqn:Whole_Sys_CGPot}
\end{equation}
where $\langle ij \rangle$ denotes a summation over neighboring cells, and $\chi$ is the neighboring class between cells $i$ and $j$.
In Eq.~\eqref{eqn:Whole_Sys_CGPot}, $H_{n_i}$ is the contribution to the total free energy of the system provided by the $n_i$ guests that, according to the occupancy configuration array $\mathbf{n}$, are located in cell $i$, whereas $K^{\chi}_{n_i,n_j}$ is the contribution provided by the effective interaction between the $n_i$ molecules in cell $i$ and the $n_j$ molecules in cell $j$, given that $i$ and $j$ are neighbors of class $\chi$.
The probability of configuration $\mathbf{n}$ to occur, $p_\mu(\mathbf{n})$, satisfies $p_\mu(\mathbf{n})\propto\exp\{-\beta\Omega_\mu(\mathbf{n})\}$, with $\beta= 1/k_B T$, where $k_B$ is the Boltzmann constant.
It is the scope of our research to find a set of $H$'s and $K$'s [see Eq.~\eqref{eqn:Whole_Sys_CGPot}] such that the coarse-grained probability distribution $p_\mu(\mathbf{n})$ matches with the probability of configuration $\mathbf{n}$ estimated from classical GCMC simulations of the FG system; a requirement that we want $H$'s and $K$'s parameters to satisfy is \emph{locality}, meaning that \emph{they would not depend on any global variable other than temperature}. 

Being $H_{n_i}$ and $K^{\chi}_{n_i,n_j}$ meant as (local) free energies, the corresponding contributions to the partition function of the system are given by 
\begin{align}
 Q_n = e^{-\beta H_n}, \qquad Z^{\chi}_{n_1,n_2} = e^{-\beta K^{\chi}_{n_1,n_2}}
\end{align}
respectively.
In order to obtain the $Q_n$ parameters, we first carry out GCMC simulations of \emph{one single cell} of the FG system at several values of $\mu$; for each one of them, we use the GCMC results to estimate the occupancy distribution $p^o_\mu(n)$, that is the probability that the cell we simulated contained exactly $n$ guest molecules.
For such one-cell system the CG potential is then
\begin{align}
    \Omega^o_\mu(n)= -\mu n + H_n
    \label{eq:Ohm0}
\end{align}
and its relation with the 
equilibrium probability of a cell to have occupancy $n$ is
$p^{o}_{\mu}(n)\propto{e^{\beta \mu n}Q_n}$.
Therefore, for any two different occupancies $n$ and $n'$ we can write
\begin{equation}
    \frac{Q_n}{Q_{n'}}= 
    \frac{
      e^{-\beta\mu n}\, p^o_\mu(n)
    }{
      e^{-\beta\mu n'}\, p^o_\mu(n')
    },
    \label{eqn:Recur_Q}
\end{equation}
and use such relation to estimate the $Q$'s recursively, starting from $H_0=0$ (or equivalently $Q_0= 1$). 
As the accuracy of each bar of the $p^o_\mu(n)$ histogram we estimated from molecular GCMC would slightly vary from one chemical potential to the other, a weighting procedure such as the one described in our previous work\cite{Pazzona2018} can be used to obtain the $\mu$-independent set of $Q$'s we are looking for.

\begin{figure}[t]
  \includegraphics[width=3.3in]{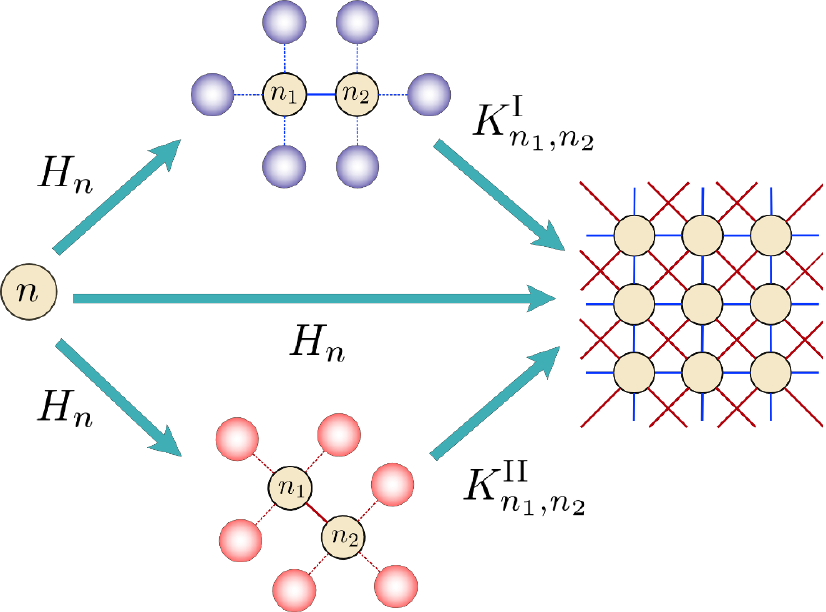}
  \caption{\footnotesize{CG workflow scheme for a square lattice with classes I and II, and $\nu^{\text{I}}=\nu^{\text{II}}=4$. On the left side: the closed single cell model, employed for the calculation of the $H_n$ parameters. In the middle: I-IPA and II-IPA models for the calculation of the pair interaction parameters (I-IPA considers only neighbors of class I, while II-IPA considers only neighbors of class II). The mean field cells are indicated with a color gradient according to the respective classes: blue for class I and red for class II. On the right side: a portion of the resulting CG lattice model where each cell is connected to neighbors of both class I and class II.}}
  \label{fgr:CG_flow}
\end{figure}

In order to estimate the $K$'s, i.e.~the pair-interaction terms, we need to employ a different model, where additional assumptions are introduced.
As different neighboring classes contribute differently to the total free energy of the system, we associate each one of them, say class $\chi$ (where $\chi=$ I or II), with its own set of probability distributions. 
Each element of such set is the bivariate occupancy distribution $p^{\chi}_{\mu}(n_1,n_2)$ computed at chemical potential $\mu$. For any two specific values of $n_1$ and $n_2$, it represents the probability that two neighboring cells of neighboring class $\chi$ contain $n_1$ and $n_2$ guests, respectively, given that the chemical potential is $\mu$. 
We estimated the histograms $p^{\chi}_{\mu}(n_1,n_2)$ from GCMC simulations of a $4 \times 4$-sized FG system where we neglected all the guest-guest interactions apart from (i) interactions between guests located in the same cell, and (ii) interactions between guests located in neighboring cells of neighboring class $\chi$, and then we establish a proper connection between the bivariate occupancy histograms $p^{\chi}_{\mu}(n_1,n_2)$ and two mean-field models within the interacting pair approximation (IPA), namely one IPA model for neighborhood class I, and another one for neighborhood class II.
Every such $\chi$-IPA dedicated model is made of one pair of explicit cells (``1'' and ``2'', respectively with occupancy $n_1$ and $n_2$; we call these cells \emph{explicit} because $n_1$ and $n_2$ are assigned well-defined integer values) that are class $\chi$ neighbors of one another, \emph{plus} $2\nu^{\chi}-2$ surrounding cells with unspecified occupancy---i.e., $\nu^{\chi}-1$ \emph{mean-field} cells interacting with cell 1, and $\nu^{\chi}-1$ more mean-field cells interacting with cell 2.
The structure of the $\chi$-IPA models and their role in the coarse-graining process is depicted in Fig.~\ref{fgr:CG_flow}.
The nature of such additional cells is \emph{mean-field} in the sense that any information about their state stay hidden inside the global variable $\mu$. We assume the guests in every such cell to interact only with the guests in either one of the two \emph{explicit} cells of the pair (namely, cell 1 \emph{or} cell 2); the effective interaction between an explicit cell of occupancy $n$ and any of its $\nu^{\chi}-1$ mean-field neighbors can be reasonably thought of as $\overline{K}^{\chi}_{\mu,n}\sim\sum_m K^{\chi}_{n,m} p^{\chi}_{\mu}(n, m)/p^{\chi}_{\mu}(n)$, with $m$ as a fictitious occupancy of the mean-field cell. Such contribution is a $\mu$-dependent mean-field term but, as we are about to show, mean-field terms will cancel out in the final formula for the pair interactions.

Given the above considerations, for each $\chi$-IPA model the total CG potential is
\begin{align}
    \notag \Omega^{\chi}_{\mu}(n_1,n_2)= & \Omega^{o}_{\mu}(n_1) + \Omega^{o}_{\mu}(n_2) + K^{\chi}_{n_1,n_2} \\
    & + (\nu^{\chi} - 1)(\overline{K}^{\chi}_{\mu,n_1} + \overline{K}^{\chi}_{\mu,n_2}),
    \label{eqn:IPA_CGPOT}
\end{align}
where the $\Omega^{o}_{\mu}$ terms are defined according to Eq.~\ref{eq:Ohm0}, and $\overline{K}^{\chi}_{\mu,n_1}$, $\overline{K}^{\chi}_{\mu,n_2}$ are mean-field interaction terms.
Now, there are two basic assumptions we rely upon in this work: 
(i) the contribution from each class to the total free energy does not depend on the contribution from any other class, and (ii) each $\chi$-IPA model is a good approximation of the reference system when \emph{only} the interactions through the $\chi$ class and the interactions inside every cell are active.
The first assumption enables us to write the CG potential for a \emph{single} cell interacting with its $\nu^{\chi}$ neighbors of class $\chi$ as 
\begin{equation}
    \Omega^{\chi}_\mu(n)= \Omega^{o}_\mu(n) + \nu^{\chi}\overline{K}^{\chi}_{\mu,n} ,
    \label{eqn:SingleCellMF_CGPot}
\end{equation}
whereas the second assumption establishes the proportionality between $\exp{[-\beta \Omega^{\chi}_{\mu}(n_1,n_2)]}$ and the $p^\chi_\mu(n_1,n_2)$, i.e.~the histogram we evaluated through GCMC simulations of the FG system.
If we consider another pair of occupancies $(n'_1,n'_2)$ for two neighboring cells of class $\chi$, we can eliminate the mean-field terms from \eqref{eqn:IPA_CGPOT} and \eqref{eqn:SingleCellMF_CGPot}, and obtain the following recurrence relation:
\begin{align}
  \notag  \frac{Z^{\chi}_{n_1,n_2}}{Z^{\chi}_{n'_1,n'_2}}= \left(\frac{e^{\beta \mu n'_1}Q_{n'_1}\,e^{\beta \mu n'_2}Q_{n'_2}}{e^{\beta \mu n_1}Q_{n_1}\,e^{\beta \mu n_2}Q_{n_2}}
\right)^{\frac{1}{\nu^{\chi}}} \\
\times \left( \frac{p^{\chi}_\mu(n'_1)\,p^{\chi}_\mu(n'_2)}{p^{\chi}_\mu(n_1)\,p^{\chi}_\mu(n_2)} \right)^{1 - \frac{1}{\nu^{\chi}}} \frac{p^{\chi}_\mu(n_1,n_2)}{p^{\chi}_\mu(n'_1,n'_2)} ,
    \label{eqn:recur_Z}
\end{align}
which starts with $Z^\chi_{0,n}=Z^\chi_{n,0}=Z^\chi_{0,0}=1$.
Eq.~\eqref{eqn:recur_Z} becomes operative once we have knowledge of all the required probability histograms---which we gain from simulations of the FG system with the proper interaction settings. 
Also in this case, the weighting procedure described in our previous work\cite{Pazzona2018} can be used to obtain a $\mu$-independent set of $Z$'s.\\

\noindent\emph{Data pre-processing at low $T$}.
According to Eqs.~\eqref{eqn:Recur_Q} and~\eqref{eqn:recur_Z}, the estimation of CG parameters relies on the occupancy histograms obtained from the GCMC simulation of the reference (FG) system, under a variety of conditions (i.e.~by excluding some or all the interactions between molecules located in different cells).
Now, GCMC simulations are finite; therefore, at each chemical potential, histogram bars in the nearness of the probability maximum will be better sampled than those far from it.
At low temperatures, the noise and the irregular shape in GCMC histograms might partly compromise the accuracy of CG results in terms of occupancy correlations in space.
In such cases we found out very effective to \emph{process} the GCMC histograms \emph{before} feeding them into the recurrence relations~\eqref{eqn:Recur_Q} and~\eqref{eqn:recur_Z}.
The ``processing'' consists in replacing the original GCMC bivariate occupancy histograms, $p^{\chi}_\mu(\cdot,\cdot)$, with new distributions, $\pi^{\chi}_\mu(\cdot,\cdot)$, whose properties should approximate a number of selected properties (namely, marginal means, marginal variances, and covariance) of the original ones, but are ``less noisy''.
We define these new distributions according to a bivariate quantized Gaussian distribution model:
\begin{align}
    \pi^{\chi}_\mu(n_1,n_2) \propto
    \exp\left[ - \frac{z}{2(1-r^2)}\right]
    ,
    \label{eqn:Fit_Mod}
\end{align}
where
\begin{align}
    z= \frac{(n_1-a_1)^2}{s_1^2}
        + \frac{(n_2-a_2)^2}{s_2^2}
        - \frac{2 r (n_1-a_1) (n_2-a_2)}{s_1 s_2}
        .
    \label{eqn:Fit_Modz}
\end{align}
In this model there are five parameters, namely $a_1$, $a_2$, $s_1$, $s_2$, and $s_{12}$ (the parameter $r$ is defined as $r= s_{12}/s_1 s_2$), but only three of them are independent, because $a_1=a_2$ and $s_1=s_2$.
This is due to the fact that the occupancies $n_1$ and $n_2$ have the same nature (i.e.~they are defined over two equivalent subvolumes), so that the two marginal averages are the same, and also the two marginal variances are the same. 
The distribution in~\eqref{eqn:Fit_Mod} is a \emph{quantized} Gaussian because variables $n_1$ and $n_2$ are integer numbers (moreover, they are defined over a finite range of non-negative values), this causing $\pi^{\chi}_\mu(\cdot,\cdot)$ to bear little to no resemblance with a (continuous) normal distribution. Therefore, in general, there is no correspondence neither between $a_1,a_2$ and the marginal means, nor between $s_1^2$, $s_2^2$ and the marginal variances, nor between $s_{12}$ and the covariance. $a_1$, $s_1$, and $s_{12}$ are rather free parameters that we direct-search optimize to produce $\pi^{\chi}_\mu(\cdot,\cdot)$ histograms that reasonably approximate the original distributions $p^{\chi}_\mu(\cdot,\cdot)$, in terms of marginal means, marginal variances, and covariance.

\section{Results and discussion}\label{Res_and_Dis}
We developed the present CG scheme considering two host-guest systems: united atom methane adsorbed (i) on a single layer of graphene, and (ii) between two graphene layers.
In the latter system the interlayer spacing is $12$~\AA.
For both systems, we performed the same partitioning, consisting in a single layer tiling of tetragonal cells with $a= 17.1$~\AA, and $c= 12$~\AA (see Fig.~\ref{fgr:Snap}).
The cut-off of pair-wise interactions was also set at $12$~\AA.
Being all the cells on the same layer, we can actually see this partitioning as a two-dimensional system of adjacent squares.
The host materials were represented as rigid structures, with each carbon atom modeled as a Lennard-Jones particle,\cite{Abbaspour2016} and each methane molecule as a single Lennard-Jones bead.\cite{Dubbeldam2005}

\begin{figure}[t!]
  \includegraphics[width=3.3in]{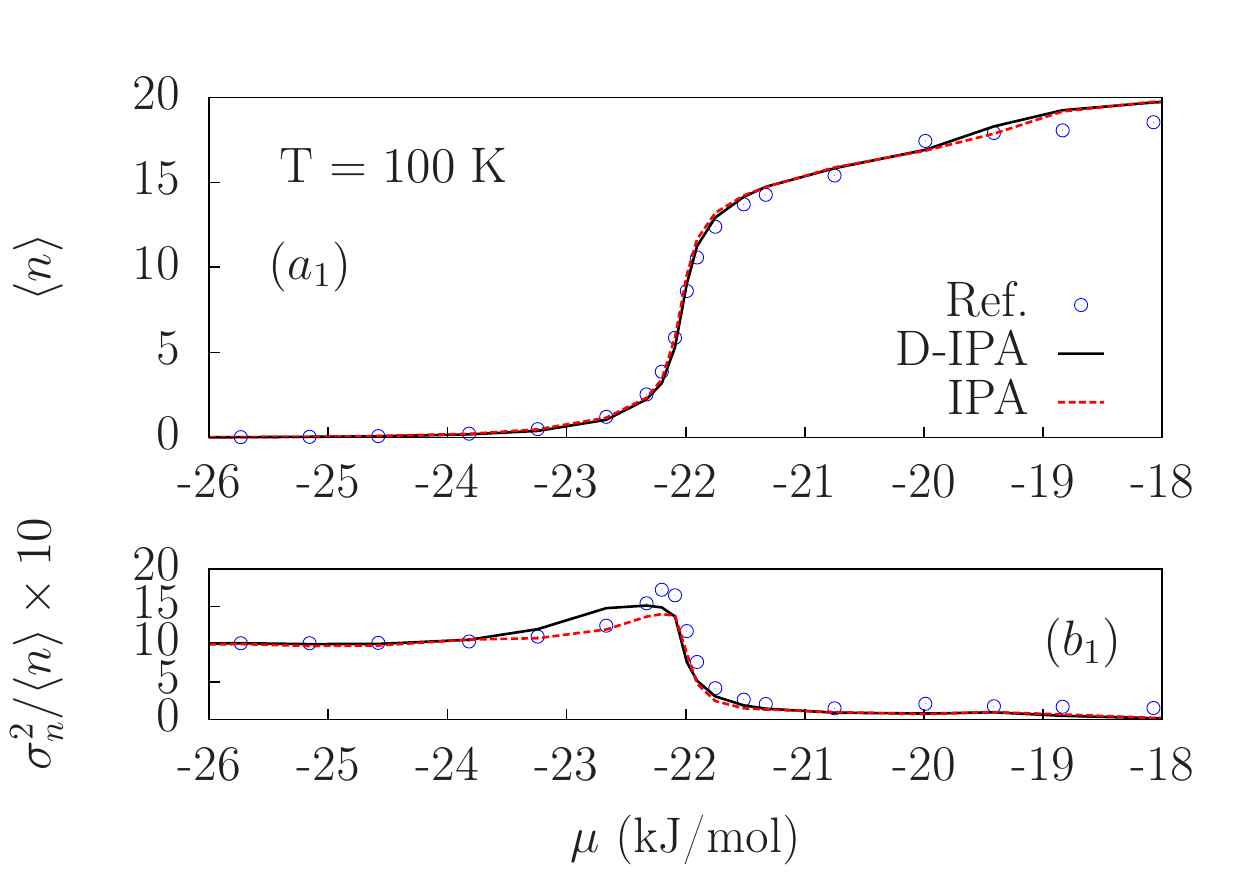}
  \includegraphics[width=3.3in]{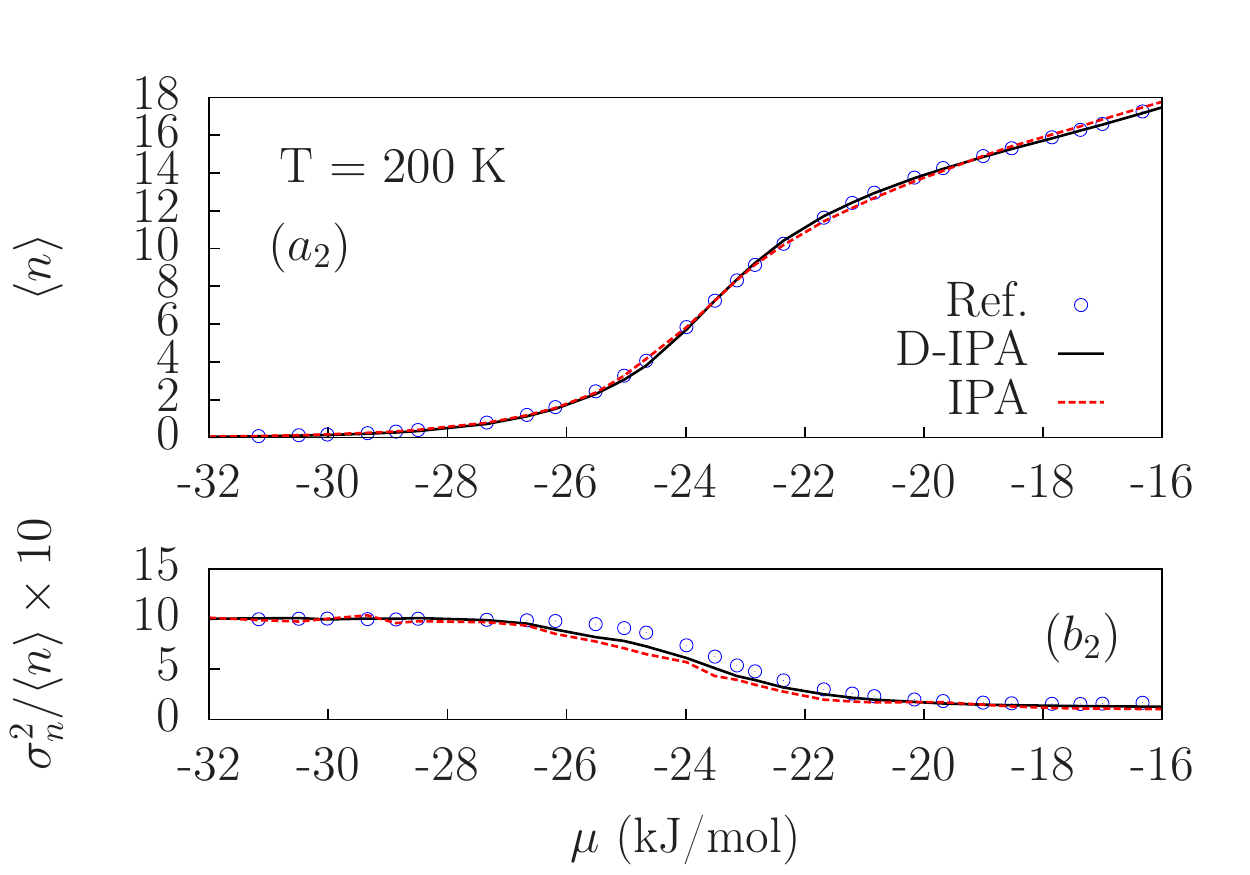}
  \includegraphics[width=3.3in]{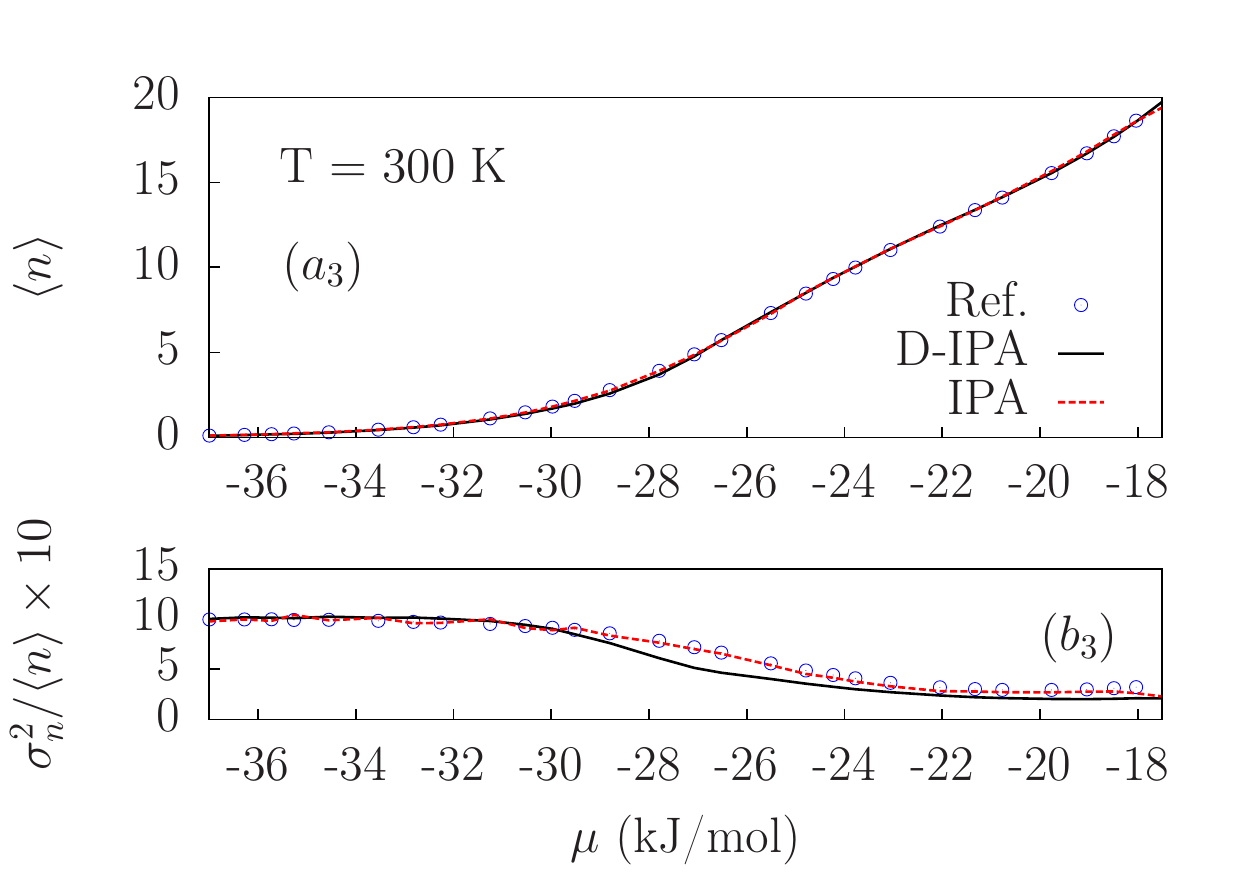}
  \caption{\footnotesize{Isotherms and reduced variance for methane on single layer graphene at temperatures $100$, $200$ and $300$ K. Isotherms are shown in subfigures labeled with letter $a$, reduced variances are shown in subfigures labeled with letter $b$. Blue empty circles represent the reference GCMC simulations with all classes active, solid black lines represent the CG simulations with data-preprocessing (D-IPA), dashed red lines represent the CG simulations without data-preprocessing.}}
  \label{fgr:Iso_Vars}
\end{figure}
Mapping such systems to the lattice model leads to a topology analogous to the King's graph, which can be imagined as the overlap of a square lattice with another square lattice rotated by a $45^o$ with respect to the first one and stretched by a factor $\sqrt{2}$ (see Fig.~\ref{fgr:CG_flow}).

For both FG systems, classical GCMC simulations were performed in a variety of different conditions in order to separate the cell-to-cell interactions, according to the prescriptions we illustrated in the description of the model.
We considered each system at three different temperatures (100, 200, and 300 K), and for each temperature we conducted a fine scan of $\mu$ values.

After calculating at each temperature the local free energy terms $H_n$ and $K^{\chi}_{n_1,n_2}$, both with and without resorting to the pre-processing of histograms, we simulated the so obtained CG lattice models in the grand canonical ensemble with the Metropolis-Hastings scheme. 
Here we compare the static properties of the FG and the corresponding CG systems in terms of adsorption isotherms, occupancy fluctuations, and occupancy covariances.
Adsorption isotherms are reported as the \emph{loading} (i.e.~average occupancy, $\langle n \rangle$) \emph{vs.}~$\mu$, whereas FG and CG fluctuations are compared in terms of the reduced variance, $\sigma^2_{n,Red}={\sigma^2_n}/{\langle n \rangle}$, where $\sigma^2_n$ is the occupancy variance for a single cell.~\cite{Hansen1986,Truskett1998}
Comparisons of spatial correlations (i.e., covariance) for each neighboring class are carried out in terms of Pearson correlation coefficients, which in the present case read $\rho^{I}= \sigma^{I}_{12} / \sigma_n^2$ and $\rho^{II}= \sigma^{II}_{12} / \sigma_n^2$ for class I and class II respectively, where $\sigma^\chi_{12}$ is the occupancy covariance of the pair occupancy distribution $p^\chi_\mu(\cdot,\cdot)$ for class $\chi$, and $\sigma_n^2$ is the marginal variance.\\

\noindent
\emph{Methane on single layer graphene.}
Results of numerical simulations are shown in Fig.~\ref{fgr:Iso_Vars}, where ``Ref'' denotes results from GCMC simulations of the FG systems, while ``IPA'' means coarse-graining without histogram pre-processing, and ``D-IPA'' indicates coarse-graining \emph{with} histogram pre-processing.
From one GCMC simulation of the FG system to the next, the chemical potential is changed by a small amount until the completion of a single layer of adsorbed methane molecules.

\begin{figure}[t!]
  \includegraphics[width=3.3in]{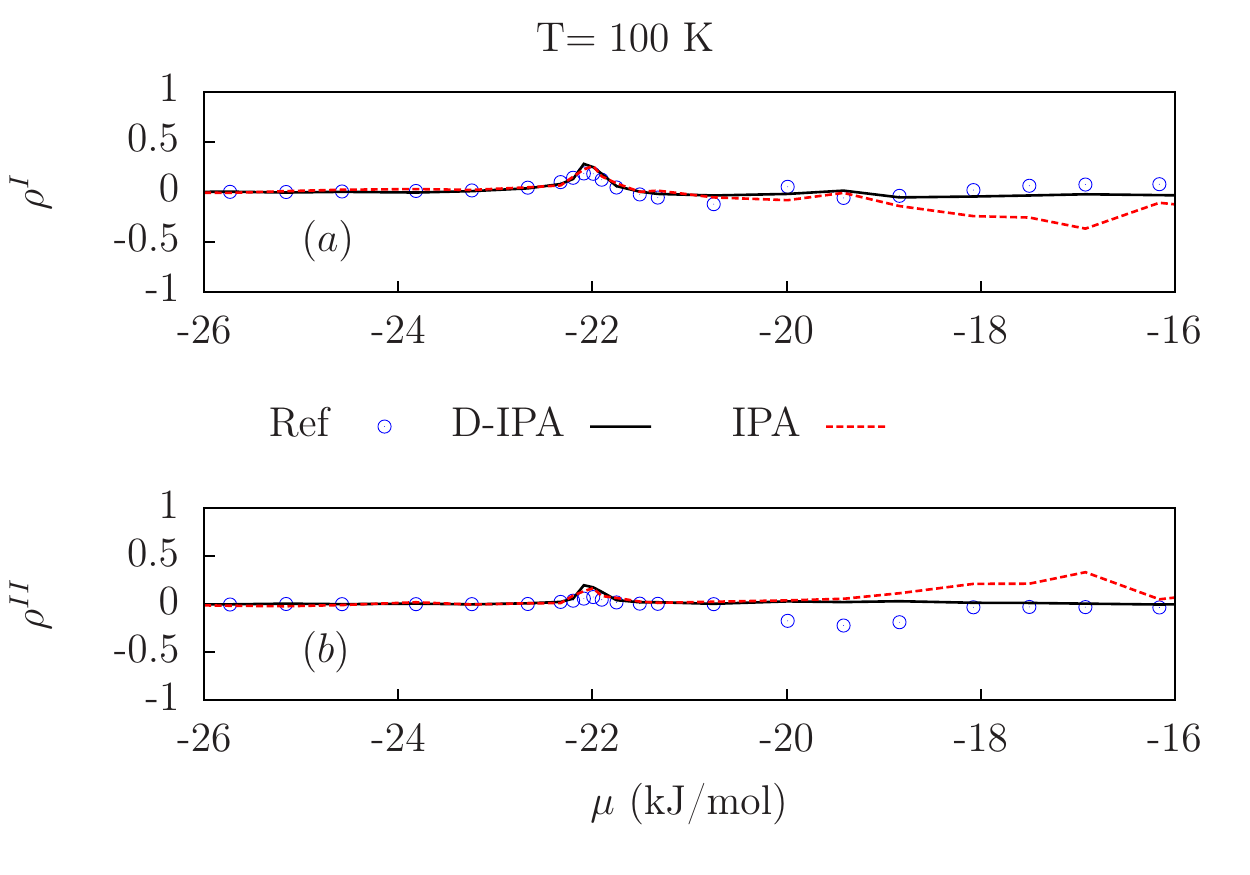}
  \caption{\footnotesize{Class-wise Pearson correlation coefficients for methane on single layer graphene at $100$ K. Results for class I are shown in the subfigure labeled with letter $a$, results for class II are shown in the subfigure labeled with letter $b$. Blue empty circles represent the reference GCMC simulations with all classes active, solid black lines represent the CG simulations with data-preprocessing (D-IPA), dashed red lines represent the CG simulations without data-preprocessing.}}
  \label{fgr:Corr}
\end{figure}
Increasing the temperature in the FG system yields a smoothing and straightening effect both on the isotherms and the occupancy fluctuations, this effect being due to the decrease of correlations between the host material and the guest molecules.
Both the IPA and the D-IPA models perform with a comparable accuracy with respect to the FG results, which is always quantitative for the isotherms and semi-quantitative for the fluctuations.
More specifically, the original isotherms are quantitatively matched at all three temperatures by both CG models, with the IPA case providing a nearly perfect match.
The situation is the same for the reduced variances and the covariances, except for the lowest temperature case (100 K) at high loadings, where an increase of correlations between neighboring cells is observed in the steep region of the isotherm ($\mu \approx -22$ kJ/mol), where we have the filling of one methane layer upon the graphene sheet (Fig.~\ref{fgr:Corr}). Such increase in correlations causes GCMC histograms to assume a very ``irregular'' shape. Noise becomes then a relevant issue during the histogram evaluation, and the recursive nature of relations~\ref{eqn:Recur_Q} and ~\ref{eqn:recur_Z} for the calculation of the free-energy contributions leads to propagation of error in the estimation of occupancy histograms. Under such conditions, pre-processing the histograms proved then to be crucial, leading the CG model back to quantitative matching.\\
\begin{figure}[t!]
  \includegraphics[width=3.3in]{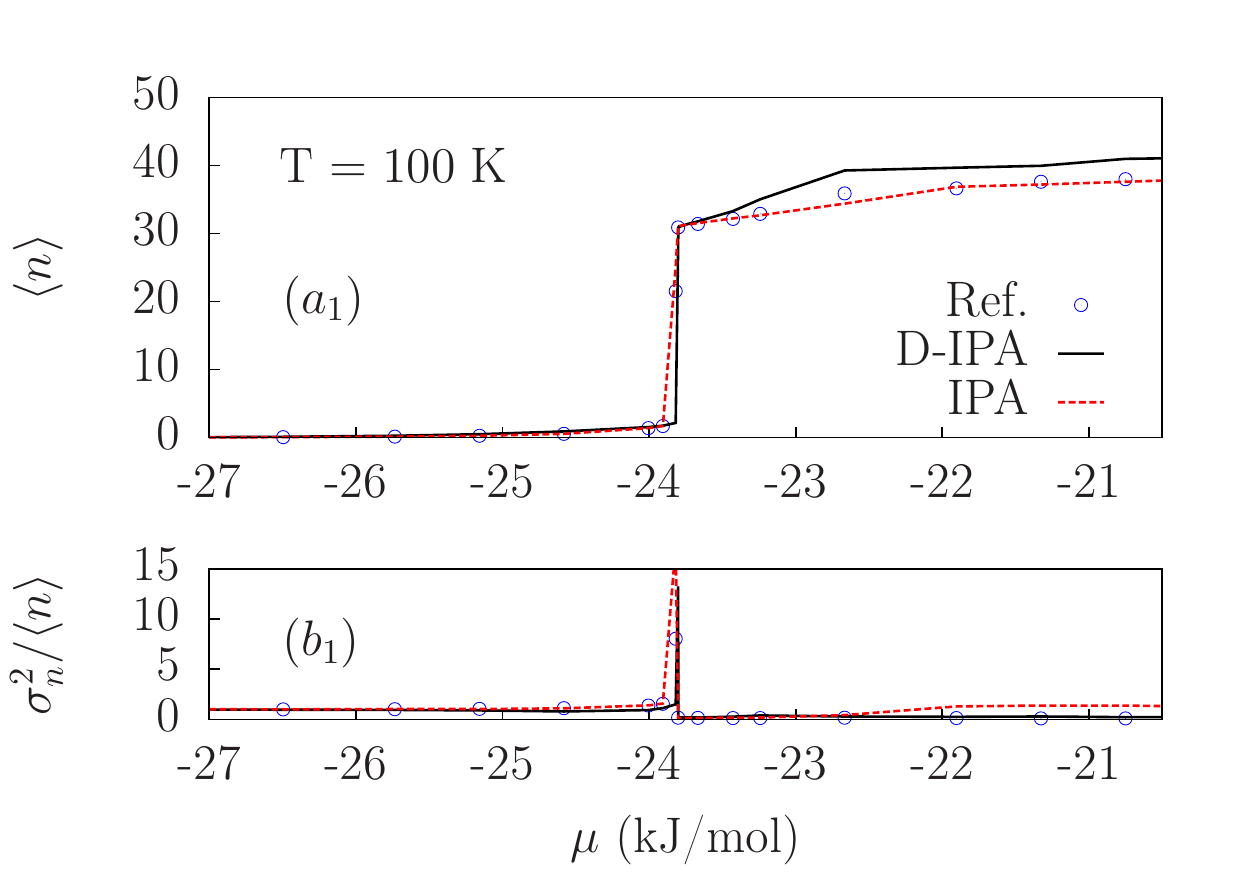}
  \includegraphics[width=3.3in]{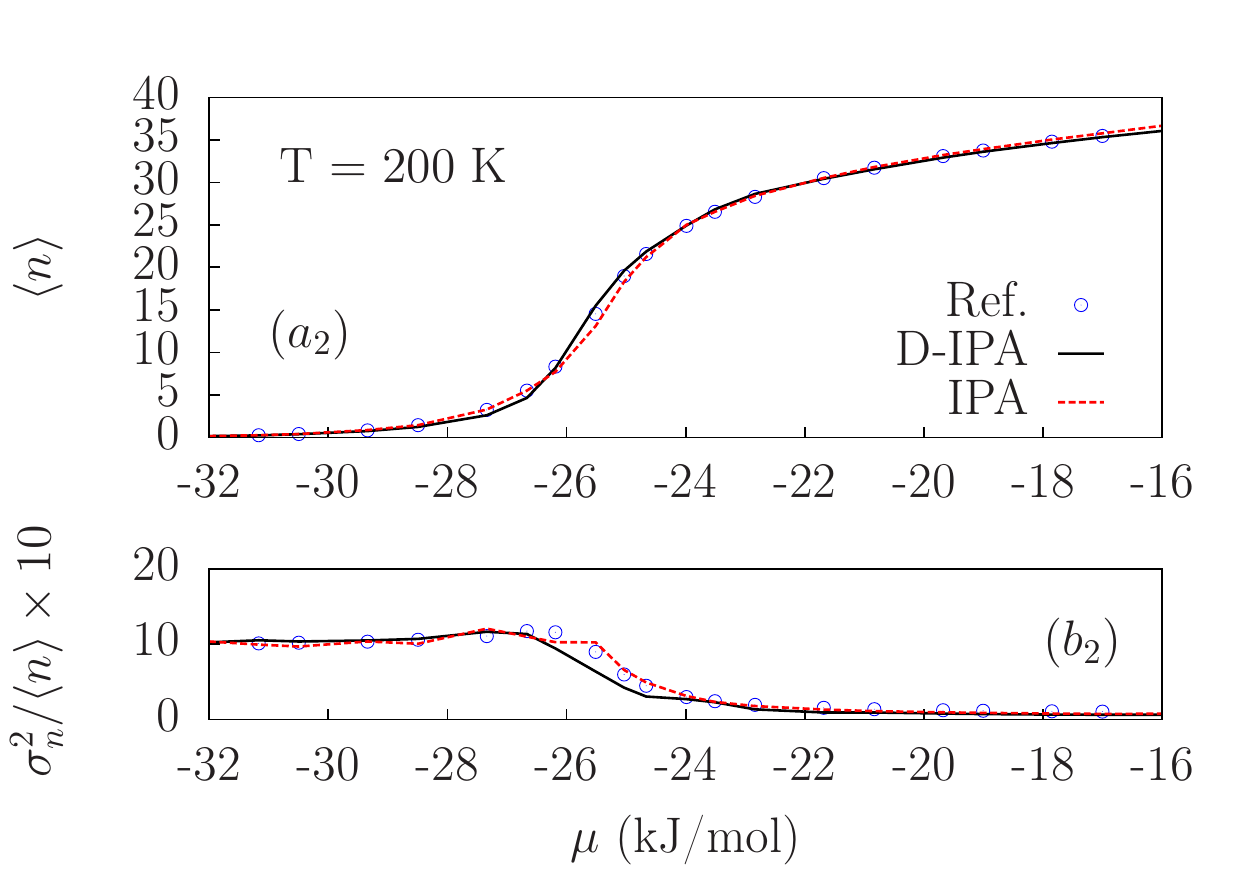}
  \includegraphics[width=3.3in]{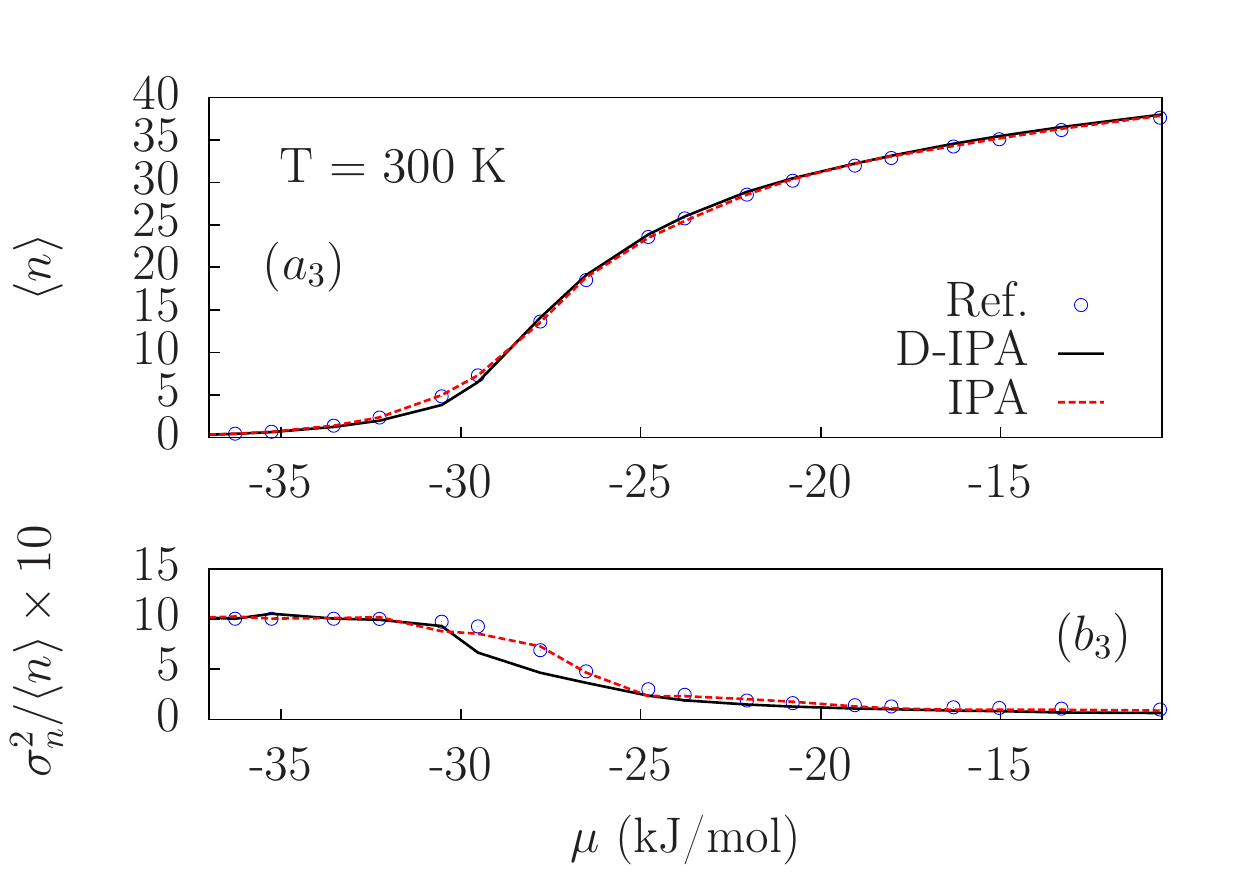}
  \caption{\footnotesize{Isotherms and reduced variance for methane between two graphene layers at temperatures $100$, $200$ and $300$ K. Isotherms are shown in subfigures labeled with letter $a$, reduced variances are shown in subfigures labeled with letter $b$. Blue empty circles represent the reference GCMC simulations with all classes active, solid black lines represent the CG simulations with data-preprocessing (D-IPA), dashed red lines represent the CG simulations without data-preprocessing.}}
  \label{fgr:Iso_Vars_BI}
\end{figure}

\noindent
\emph{Methane between two graphene layers.}
In this case, GCMC simulations of the FG system were conducted within a chemical potential range which allows for the filling of a double layer of methane molecules in the interlayer space (that amounts to 12~\AA). 
The FG and CG results (adsorption isotherms and reduced variances) for this system are shown in Fig.~\ref{fgr:Iso_Vars_BI}.
The accuracy scenario of the CG representations is comparable to the one obtained for the previous system, with quantitative agreement attained in all but the lowest temperature/high loading case.

\begin{figure}[t!]
  \includegraphics[width=3.3in]{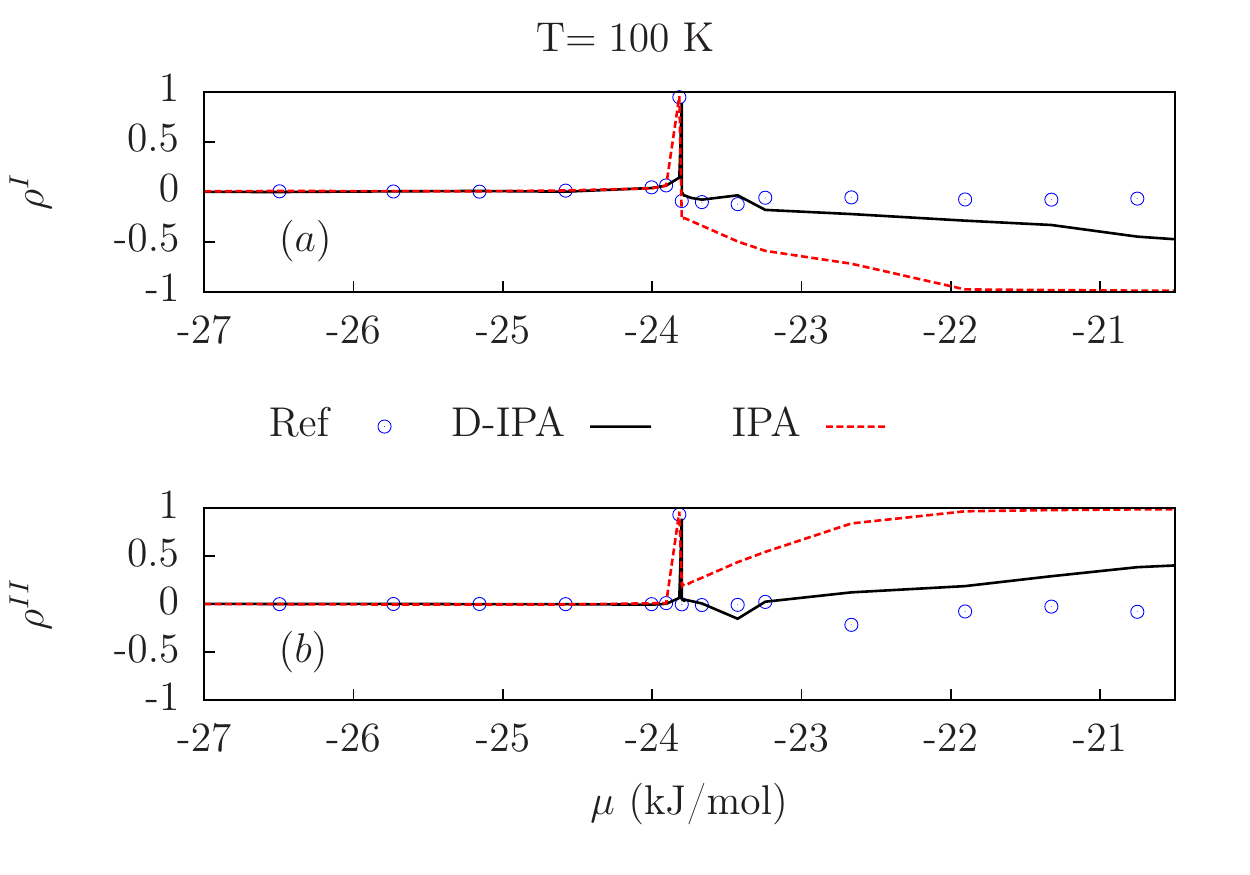}
  \caption{\footnotesize{Class-wise Pearson correlation coefficients for methane and double layer graphene at $100$ K. Results for class I are shown in the subfigure labeled with letter $a$, results for class II are shown in the subfigure labeled with letter $b$. Blue empty circles represent the reference GCMC simulations with all classes active, solid black lines represent the CG simulations with data-preprocessing (D-IPA), dashed red lines represent the CG simulations without data-preprocessing.}}
  \label{fgr:CorrBI}
\end{figure}

A major difference between this and the single-layer case lies in the steepness in the step in the adsorption isotherm, which for the double graphene layer case at $T=100$ K is observed at $\mu\sim -24$ kJ/mol, and is definitely abrupt: a chemical potential increase of about 0.2 kJ/mol causes the loading to sharply rise from 1.4 to 31 guest molecules per cell---correspondingly, the reduced variance shows a sharp peak.
From a molecular point of view, this corresponds to the sudden and simultaneous formation of two adsorbed methane layers between the two graphene sheets.
A detailed molecular-level analysis of this transition falls beyond the scope of this work, i.e.~the production of a CG model that could effectively reproduce also a behavior like this one, and will be the subject of further investigations.
Also in this case, however, the pre-processing (D-IPA curves in Fig.~\ref{fgr:CorrBI}) allowed for the production of a set of CG interaction parameters that significantly improved the agreement in terms of spatial correlations at high loadings. 
By looking at the D-IPA curves in Fig.~\ref{fgr:CorrBI}(a1) for $\mu > -24$ kJ/mol, we can see that such improvement comes along with an improvement in the single-cell reduced variance as well, but also with a slight accuracy loss in the adsorption isotherm---which could be made even slighter, but at the considerable cost of increasing the complexity of the coarse-graining model, e.g.~by including a further CG equation [besides Eqs.~\eqref{eq:Ohm0} and~\eqref{eqn:IPA_CGPOT}] describing three-term interactions.
Therefore, we believe that the accuracy in the adsorption isotherm can be still considered very satisfactory, despite the class-independence assumption we made in order to keep the CG model definition as simple as possible.\\

The histogram pre-processing improved the correlations in situations in which the original pair-occupancy histograms obtained through GCMC were certainly affected by non-negligible accuracy issues. 
In fact, at low temperature and high density (but not close to the adsorption step) the occupancy fluctuations are low; correspondingly, the occupancy distributions turn out to be sharply peaked. Now, the cell occupancy varies within a relatively small range, which goes up to about 20 and 40 molecules per cell, respectively for the case of methane in a single graphene layer and within a double graphene layer. As a consequence, occupancy histograms being sharply peaked imply good sampling of only a limited number of occupancy pairs, namely, those that are very close to the average value. Any other occupancy pair is sampled poorly. Eq.~\eqref{eqn:IPA_CGPOT}, i.e.~the one that contains information about class-wise occupancy correlations, is the CG equation that is most seriously affected by such accuracy, and the diverging correlations shown in Figs.~\ref{fgr:Corr} and~\ref{fgr:CorrBI} are the end result. 
In the vicinity of the adsorption step the situation is even more complicated: the variances are \emph{very high}, but this does not necessarily imply that the corresponding distributions are short and wide---more generally, the occupancy distributions under such conditions are no longer unimodal, and can not be considered stable (i.e., very small changes in $\mu$ would cause large changes in the shape of distributions).
When facing such problems, the first solution that comes to mind would be to carry out much longer simulations, in order to have significantly more data to take into account while estimating the occupancy histograms. However, we wanted to find out how much the CG model could be improved with just the input data we had, without adding more data to the source set of histograms; this is the reason why we preferred to manipulate that set by means of a ``histogram imitation technique'', rather than to perform longer GCMC runs. Of course replacing the original distributions with "fake, but better-behaving ones" means to coarse-grain a system that differs from the original one in some aspects. Nevertheless, if the CG model we want to build from some FG reference system aims to correctly imitate its occupancy correlations in space, such an operation appears legitimate.

\section{Conclusions \label{sec:Conclusions}}
We performed the spatial coarse-graining of the static occupancy-related properties of two adsorption systems, namely one and two graphene sheets with methane as the adsorbate, at various temperatures.
In order to accomplish this task, we extended the interacting-pair approximation (IPA) method\cite{Pazzona2018}, a local occupancy-based spatial partitioning approach to the coarse-graining of host-guest systems, to the case in which every subvolume of the partition is surrounded by neighboring subvolumes of two kinds.
The resulting two different kinds of spatial correlations were reproduced by local, class-wise mutual interaction parameters, defined on the basis of pair-occupancy histograms evaluated from properly tailored fine-grained GCMC simulations within a wide range of chemical potentials, $\mu$---namely, from zero-loading to the complete filling of the graphene sheet(s) with sorbate molecules.
The coarse-grained (CG) potentials we obtain are functions of the local occupancies and are temperature-dependent, but do \emph{not} depend on any other global variable (such as, e.g., overall density or chemical potental); this enables us to use the same set of CG potentials at any value of $\mu$ within the range of interest.
We evaluated the quality of coarse-graining in terms of agreement between the properties of the local occupancy distributions of the coarse-grained (CG) systems, and the properties of the same distributions for the corresponding reference, fine grained (FG) systems.
The results showed a very satisfactory agreement in almost all the scenarios we investigated. Only at low temperature (100 K) and high densitiy both systems required a pre-processing of the pair-occupancy histograms over which the CG potentials are defined, in order to allow for the production of realistic CG correlations despite the relatively poor accuracy with which they were sampled, without resorting to longer sampling runs.
This pre-processing prescribed the replacement of the original GCMC histograms with quantized Gaussian distributions with similar means, variances and covariance; the improvement we obtained from it was especially relevant for the double-layer case at 100 K, where the adsorption isotherm shows an abrupt and steep loading change at intermediate loadings---a scenario where accuracy issues in the source GCMC histograms may prevent the CG parameters from producing correct occupancy correlations at high loading.




\end{document}